\documentclass[final,5p,times,twocolumn]{elsarticle}

\usepackage{graphicx}

\usepackage{amssymb}

\usepackage{lineno}
\usepackage{caption}
\usepackage{supertabular}
\usepackage{longtable}
\usepackage{amsmath,amssymb,amsfonts}
\usepackage{algorithmic}
\usepackage{graphicx}
\usepackage{textcomp}
\usepackage{xcolor}
\usepackage{tabularx,booktabs}
\usepackage{pgfplots}
\usepackage{url}
\usepackage{hyperref}
\usepackage{mdframed}

\newmdenv[
  topline=false,
  bottomline=false,
  rightline=false,
  skipabove=\topsep,
  skipbelow=\topsep
]{siderules}

\newcolumntype{L}{>{\raggedleft\arraybackslash}X}
\newcolumntype{R}{>{\raggedright\arraybackslash}X}

\journal{Journal of Systems and Software}

\begin{document}

\begin{frontmatter}

\title{On Misbehaviour and Fault Tolerance in Machine Learning Systems}

\author{Lalli Myllyaho}
\author{Mikko Raatikainen}
\author{Tomi Männistö}
\author{Jukka K. Nurminen}
\author{Tommi Mikkonen}
\address{University of Helsinki}

\begin{abstract}
Machine learning (ML) provides us with numerous opportunities, allowing ML systems to adapt to new situations and contexts. At the same time, this adaptability raises uncertainties concerning the run-time product quality or dependability, such as reliability and security, of these systems. Systems can be tested and monitored, but this does not provide protection against faults and failures in adapted ML systems themselves. We studied software designs that aim at introducing fault tolerance in ML systems so that possible problems in ML components of the systems can be avoided. The research was conducted as a case study, and its data was collected through five semi-structured interviews with experienced software architects. We present a conceptualisation of the misbehaviour of ML systems, the perceived role of fault tolerance, and the designs used. Common patterns to incorporating ML components in design in a fault tolerant fashion have started to emerge.  ML models are, for example, guarded by monitoring the inputs and their distribution, and enforcing business rules on acceptable outputs. Multiple, specialised ML models are used to adapt to the variations and changes in the surrounding world, and simpler fall-over techniques like default outputs are put in place to have systems up and running in the face of problems. However, the general role of these patterns is not widely acknowledged. This is mainly due to the relative immaturity of using ML as part of a complete software system: the field still lacks established frameworks and practices beyond training to implement, operate, and maintain the software that utilises ML. ML software engineering needs further analysis and development on all fronts.
\end{abstract}


\end{frontmatter}

\section{Introduction}

Machine learning (ML) has pierced through the society. A ML system aims at improving its behaviour through experience~\cite{mitchell2006discipline}. Essentially, this means that a program learns to give correct outputs for given inputs without explicitly being programmed to do so. Learning algorithm and large amounts of training data are used to tune a statistical model that represents the problem field. Its ability to adapt and predict situations its developers never thought of presents unprecedented possibilities. However, this adaptability  raises questions concerning the run-time quality or dependability~\cite{ieeedependability}, such as reliability and security, of systems utilising ML: Can we trust the systems? How? Not only does the adaptability propose potential problems but a great deal of the time, the used ML models make a right prediction, for example, 99 \% of the time \cite{ramanathan2016integrating}. The percentage itself is impressive but what to do with the remaining 1\%? And where do these inputs lie? In addition, an ML model often keeps on learning after deployment, possibly distancing itself from the initial tested ML model \cite{amershi2019software}, leaving new features untested.

Research on ML dependability has mostly focused on testing  ML models, and less on the whole systems that utilize ML \cite{zhang2020machine}. The latter includes how the systems should be monitored once ML models have been deployed. Along with testing, verification, and validation of systems, the methods applied upon and after deployment of the ML systems are vital to ensure the systems remain dependable. Architectural designs have been suggested to protect the systems from hardware failures and malicious attacks (e.g. \cite{kriebel2018robustness}), but there is little emphasis on architectural software design to answer the inherent unpredictability and uncertainty of the utilized ML itself.

One step towards achieving dependability is fault tolerance. Traditionally, software faults have been seen as the results of design errors \cite{randell1975system}. However, due to their statistical, data-driven nature, ML systems can be seen as inherently faulty not by design, but by paradigm. Thus, unpredictable errors will emerge from deployed ML systems that cannot be captured by traditional fault-tolerance models. The question remains of how to build ML systems that detect these errors and prevent them from propagating.

However, reports on empirical experience are rare in the research literature related to fault tolerance in ML systems \cite{zhang2020machine}. As such, in this paper, the goal is to gather additional knowledge on fault tolerance solutions and beyond, and the practical applicability and reasoning of the solutions -- which are used and considered useful. We reached out to experienced software architects familiar with ML through their work. In this way, we aim to shed light on which design solutions are seen as useful by experts, which are not, and which need additional studying, thus answering the lack of research on the functionality of deployed ML models identified by Zhang \textit{et al.} \cite{zhang2020machine}.

The applied research method is the case study method~\cite{Yin2009, runeson2009guidelines}.
The data was collected through five \textit{semi-structured interviews}. The respondents were asked about their experiences with the fault-tolerant design of ML systems in general, and their experiences with an initial set of design solutions found in the ML research literature or presented in materials on fault tolerance in traditional software. This initial set is presented in more detail in Section~\ref{sec:propositions}. 

The results show that there is much to desire in the dependability of ML systems. Some patterns for fault tolerance are used in practice, but the developers and buyers often lack knowledge and frameworks to apply them. Thus, the role of fault tolerance is -- at least today -- very limited and vague in practice. This relative immaturity is not limited to fault tolerance, however, but also other phases of managing the life-cycle of ML systems. To our knowledge, this is the first attempt to gather information about fault tolerance for ML systems in one place, thus forming the basis for further research. Practitioners can use the gathered information to design more dependable ML systems.

This paper is organised as follows: Section~\ref{sec:oikeaBackground} describes the key concepts for fault tolerance, and the previous work on fault tolerance and the dependability of ML systems. Section~\ref{sec:propositions} describes the fault-tolerance solution proposals presented in literature and studied empirically in this study. The case study research method details are given in Section~\ref{sec:method}. The results of the interviews are described in Section~\ref{sec:results}.  Section~\ref{sec:discussion} and  Section~\ref{sec:validity} discuss the results and study validity. Conclusions are drawn in  Section~\ref{sec:conclusions}.

\section{Background}\label{sec:oikeaBackground}

\subsection{Dependability, faults, and ML systems}
\label{sec:background}

System \textit{dependability} means a system's trustworthiness \cite{ieeedependability}. Dependability is assessed by evaluating a system's reliability, availability, and maintainability. Sometimes additional quality characteristics, such as safety and integrity, are applied \cite{avizienis2004basic}. In other words, a dependable system -- at the very least -- delivers correct service consistently, does not suffer from long periods of down-time, and is easily corrected and altered.

Threats to dependability originate from \textit{failures}, \textit{errors}, and \textit{faults} \cite{avizienis2004basic}. Failures are deviations from the desired service. Failures result from propagating errors, i.e., incorrect functioning of the system. Errors are caused by faults that are defects in system's components (software or hardware), activated by given inputs in a given state.

Dependability can be reached by diminishing these threats or by justifying that the remaining threats in the system do no unbearable harm \cite{avizienis2004basic}. Two means of diminishing threats are \textit{fault prevention} and \textit{fault tolerance}. Fault prevention aims at not introducing faults into systems, whereas fault tolerance aims at a system design such that occurring errors are stopped from propagating and causing system failures.

As ML models are non-deterministic statistical approximations by their nature, they are bound to function completely correctly only a certain portion of the time \cite{ramanathan2016integrating}. Thus, fault prevention -- essentially building a more accurate ML model -- works only to a degree. This is true even if there are no design flaws present in the software. This is noteworthy, as design flaws are seen as the main source of faults in traditional software \cite{randell1975system}.

As faults are inherently possible or even present in ML models, preventing ML system failures cannot rely solely on fault prevention, but dependable ML systems also require fault tolerance. In software systems, fault tolerance is achieved by error detection and error recovery \cite{knight2012fundamentals}. Essentially this means that activating faults must be noticed and the resulting errors must be stopped from propagating.

Research on fault tolerance has been carried out more on traditional systems rather than those with a ML model. This is insufficient, as the deployment of ML models into software systems introduces challenges  beyond the shortcomings of the initial model~\cite{lwakatare2020devops}. In a typical project, a new ML model needs to be developed and tested. At the same time, all the other software of the system needs to be developed and operated just like in traditional software projects. However, the system is not just about the functioning of the ML model and the software around it but also about them working together as a ML-based system that brings special characteristics. System or integration testing cannot be carried out satisfactorily before the ML model is deployed. To test the system sufficiently, the developers need to assess how the ML model's requirements and evaluation affect the requirements and evaluation of the entire system. Beyond the initial version, the ML model is usually not only trained and deployed once but needs to be retrained with new data, possibly gathered by the live system, while there is still no real way of knowing whether or not the retrained model is functioning adequately before it is deployed to the system, either. Thus, updating the system may introduce a new version of the code, a new version of the data, and a new version of the model, all of which may introduce faults, and at least one of which one cannot really be sure of beforehand. Compare this to traditional software, where updating only introduces a new version of the code, and should not contain components as whimsical as ML models in the first place. The solution principles themselves are nothing new: ensuring the continuation of correct functioning is seen as part of the life-cycle of ML software \cite{zhang2020machine, breck2017ml}. However, there has been more interest in the correct functionality of the initial ML model than in making sure it keeps delivering correct service during its operational time \cite{zhang2020machine}.

\subsection{Previous work} \label{relatedWork}

ML has been shown to be problematic when applied in practical systems, and dependability is no exception. In their literature review on software engineering challenges for ML systems, Kumeno \cite{kumeno2019sofware} found that software maintenance and software quality (including run-time monitoring, fault detection, fault correction, and fault elimination, among other things) are amongst the often-reported challenges around ML faced in the software engineering literature. Thus, the problem is well-recognised in the field.

There have been efforts to grasp this problem. In their literature review, Zhang et al. \cite{zhang2020machine} take a deep dive into the testing of ML in the research literature. In their work, they separate ML testing into offline and online testing. Offline testing is basically ML model validation \cite{Wang2013}, whereas online testing includes the initial testing after model deployment, and the measures taken to ensure correct functionality beyond initial tests, such as monitoring and other fault tolerant patterns. However, the papers yielded by their search presented mostly offline testing, and very little online testing. For this reason, they point out that more research is needed on the elements of online testing. As the term online testing also includes the initial testing after deployment, to avoid confusion, we refer to post-deployment measures as a whole as \emph{continuous validation}, when not specifically talking about fault tolerance.

We are aware of studies, which include solutions what we understand as continuous validation in their system description. Some specifically mention some software fault-tolerance measures. The measures include \emph{input checker} \cite{jonsson2012towards}, \emph{output checker} \cite{prado2018machine, li2017game}, \emph{replication} \cite{yan2018design}, \emph{design diversity} \cite{fu2019retargetable}, and \emph{voting} \cite{wang2018perception}. However, in many of these papers, the details of the measures are not described very rigorously. The efficiency, implementation, or even need for them may not be reported. Even their very existence may only be revealed in a sentence, never to be returned to. This leaves room for more descriptive empirical research on the role of fault tolerance in ML systems.

All of the mentioned patterns have their counterparts in the fault-tolerant design of traditional software (cf. Knight \cite{knight2012fundamentals}). As fault tolerance has an established role in traditional software, it might be reasonable to assume that practitioners could have taken even more inspiration from the patterns used in traditional software. For this reason, additional solution proposals are taken from the realm of traditional software, specifically Knight's extensive report \cite{knight2012fundamentals}.

In addition, there has been interest in node activation observing as a testing approach for ML models, more specifically artificial neural networks (NN) \cite{pei2017deepxplore, tian2018deeptest, xie2019deephunter}. As the results indicate that different kinds of inputs tend to activate different nodes \cite{tian2018deeptest, xie2019deephunter}, it could be the case that unexpected activations could not only be used in testing but could also be used to indicate invalid inputs or untrustworthy results when the model is deployed.

This overview of previous work led us to summarise the solution proposals as described in more detail below in Section~\ref{sec:propositions}.

\section{Study propositions: Fault tolerance solutions} \label{sec:propositions}

The case study propositions~\cite{Yin2009} represented in this section direct the attention to fault tolerance solutions that are then examined within the scope of this study.
That is, since some solutions -- or \textit{patterns} -- have been mentioned in the literature (see Section~\ref{relatedWork}), and there is a tradition for fault tolerance in traditional software, we aim then in the case study to gather more in-depth empirical insight into how these solutions are perceived by software architects working on ML systems.

\subsection{Solution proposals selection}

The patterns chosen are either mentioned in earlier research in the context of ML, presented in materials for traditional software, or are a modification of some of these solutions which we perceived to be interesting (cf. Section~\ref{relatedWork}). To be included in the propositions, one of two criteria had to be met:
\begin{itemize}
    \item [1] A paper mentioned using the said pattern in a ML system, or
    \item [2] The pattern is used for traditional software, and we can come up with a scenario in which the pattern could potentially detect errors, or stop them from propagating in a ML system.
\end{itemize} In other words, the solutions were chosen based on our perception of whether or not the mentioned solutions could be seen as fault-tolerance design and whether or not we thought they could potentially be interesting in the context of ML. For example, replication (see below) was included as a proposition based on criterion 1, as it was briefly mentioned being used in a paper, whereas very similar N-copy voting \cite{knight2012fundamentals} was left out, as we could not really see how the same model voting with itself would provide any additional value for a ML system. The exception to this is activation observing, which arises from the inner workings of neural networks, but is only hinted at in earlier research \cite{tian2018deeptest}.
The solutions discussed in this paper are described below \textit{as they were initially presented}. This means that modifications and additional suggestions made to them by the respondents of the case study are presented later. Likewise, the case study is not limited to these proposals, but the respondents were encouraged to describe their experiences first -- before the propositions were presented to them -- to supplement the set, and also to reject propositions they did not feel had additional value. Thus, the list presented in this section should not be viewed as complete.

\subsection{Fault-tolerance solution proposals}

\textit{Input checker} (used by Jonsson \textit{et al.}~\cite{jonsson2012towards}) is a component that aims to prohibit such inputs from entering the ML model that could activate the ML model's faults. Thus, the faults are tolerated by limiting the potential situations in which they could cause errors. This can be done by strict limitations to the environment in which the system is deployed (as in \cite{shadrin2019designing}) but, as we are interested in software design, we emphasise the solutions in which the potentially unwanted inputs are recognised on the software level. Limitations can be strict sets or values as in the paper by Jonsson \textit{et al.}~\cite{jonsson2012towards}, but in addition to this, we include an input checker that aims to recognise inputs which are not similar to anything the ML model has seen before, henceforth referred to as \textit{novelty inputs}.

\textit{Output checker} (used by Prado \textit{et al.}~\cite{prado2018machine} and Li \textit{et al.}~\cite{li2017game}, also known as acceptance test \cite{knight2012fundamentals}) is a component which detects errors by assessing ML model's outputs and prevents errors from propagating further into other parts of the system. For example, Prado \textit{et al.} \cite{prado2018machine} set the software to limit the speed and steering angle of an electric vehicle to certain maxima, no matter what the ML-based controller output. In addition to this hard limit approach, we include whether errors of a ML model could be detected by comparing the outputs to historical data or by using another ML model, trained to detect unexpected behaviour of the primary ML model.

\textit{Model observing} means measures taken to assess the inner workings of a ML model during the operation time. Thus, the ML model's trustworthiness is based on whether or not it acts unexpectedly, regardless of its outputs' perceived correctness. This covers timer watchdogs \cite{knight2012fundamentals}: the watchdog barks if the computations done by the ML suddenly take significantly more or less time. However, we expand the idea of a timer watchdog to include other resources, such as CPU usage, and call these collectively \textit{resource consumption}. Also, we suggest an activation observer (e.g. another ML model trained for the task) to watch if the input given to the ML model activates unexpected nodes. This could be, for example, a node which has always before output small numbers and suddenly outputs a considerable number, or activation of a node which has usually only activated in relation to certain input features.

\textit{Redundancy} is a collection of approaches which rely on multiple implementations of the same component, which would be redundant if fault-freedom of the first implementation could be assured \cite{knight2012fundamentals}. We present different forms of redundancy as follows.

The first form of redundancy is \textit{recovery blocks} \cite{knight2012fundamentals}. This is organised so that if the first ML model is detected to be erroneous, the input is passed on to another ML model. If the other ML model is also detected to be erroneous, the input can be passed on to a third one and so on as long as there are more redundant ML models to pass the input on to. Recovery blocks can be set up by \textit{replication} \cite{knight2012fundamentals, yan2018design} or \textit{design diversity} \cite{knight2012fundamentals, fu2019retargetable}. In replication, the software components -- in our case, the ML model -- to which the inputs are passed on are other instances of the initial ML model; in other words, its replicas. The system was not described in great detail -- rather mentioned than described -- nor could we really come up with appropriate scenarios for it. Nevertheless, it was included based on criterion 1 so that experienced practitioners could potentially enlighten us on the situation. In design diversity, the secondary ML models, to which the input is passed on if the primary ML model is erroneous, are not instances of the primary ML model but different ML models altogether, trained for the same task. The secondary models could, for example, be a less refined, yet more robust or tested version, a simpler version dedicated to keeping the system alive, or just a plainly different model, the shortcomings of which hopefully do not overlap with the primary model. All these might risk the secondary model sharing the same defects, or even having more broad ones. We hope to find answers to this problem as well. We also considered whether or not the primary ML model should be disabled and a new primary ML model should be chosen amongst the redundant ML models if the primary ML model is detected to be erroneous.

Another form of redundancy is \textit{voting} (briefly mentioned as being used by Wang \textit{et al.} \cite{wang2018perception}, also known as N-version systems \cite{knight2012fundamentals}). In voting, different ML models decide the outcome together. Thus, the errors of a single or a few ML models can be masked, as long as the majority of the ML models are functioning correctly. Voting relies on design diversity, as instances of the same ML model can be expected to give similar outputs in all other cases than in the case of malformation. Knight \cite{knight2012fundamentals} mentions majority voting, median value, middle value, and average value as examples of heuristics that can be used as the decider for which value should be chosen. The voting could be organised by the data scientists applying ensemble learning, but we did not want to rule out less strategic approaches a priori, but rather hear what the practitioners said. For example, the models participating in the vote could include an orchestra of models used in practice over time, or they could include -- or even consist of -- off-the-shelf-models acquired elsewhere.

As voting and the divergent form of recovery blocks rely on design diversity, we also take into account how this diversity can be achieved with ML models. In traditional software, diversity has been implemented by using different development teams, development tools, design techniques, and older versions of the software \cite{knight2012fundamentals}. Gong et al. \cite{gong2019diversity} found that in prior research, diversity between different ML models has been encouraged by using different subsets of training data for each ML model, by enforcing diversity through additional regularization metrics during training, and by training multiple ML models for the same task and choosing a number of the top performing ones. However, they do not disclose how the data was gathered, which raises questions about the coverage of the work, and they approach the issue from the perspective of ML model performance rather than that of fault tolerance. Thus, we are interested in the matter not only to unveil the state-of-practice and preference of these methods, but also to find out whether or not there are more suitable methods to ensure the desired level of diversity -- and what the desired level is.

\section{Methodology}
\label{sec:method}

The research methodology of this paper is a case study~\cite{Yin2009,runeson2009guidelines}. That is, the phenomenon is studied in its real context. The chosen approach for data collection is an interview study: We interviewed multiple respondents about their experiences on several industrial projects in an exploratory manner.
\vspace{-0.25cm}
\subsection{Research goal and questions}
\vspace{-0.25cm}
As discussed in Sections~\ref{sec:background}~and~\ref{relatedWork}, there is a justified concern about the misbehaviour of ML systems. How the inherent problems actually manifest and are handled after deployment is, however, a less prominent subject in the research literature. There is evidence that some fault tolerance is present in the research literature, but the details and actual implications the patterns have on the systems are often vague. Our research goal is to capture what fault tolerance patterns are used or considered useful. Thus, we reached out to practitioners with experience on the matter to gather information both on how the problems manifest, and how they are handled. In this paper, we seek answers to the following questions:

\begin{itemize}
    \vspace{-0.25cm} \item RQ1: What kind of misbehaviour occurs in ML systems, originating from the ML model?
    \vspace{-0.25cm} \item RQ2: What is the role of fault tolerance as a means to diminish this misbehaviour?
    \vspace{-0.25cm}\item RQ3: What design patterns are useful to build fault-tolerant ML software?
\end{itemize}

\begin{table*}[t]
\footnotesize
\small
\caption{The experiences of the respondents.}
\label{tab:respondents} \centering
																	
	\begin{tabularx}{\textwidth}{>{\centering}p{1.3cm} p{3.8cm} R}
	
	\toprule

	\textbf{Respondent} & \textbf{Industrial experience in ML}& \textbf{Application domains}\\
	\midrule
	
	1
	& 20 years
	& A few media houses, Telecommunication, Banking, Consulting.\\
	
	2
	& 5 years
	& Product consulting, Education.\\
	
	3
	& 15 years
	& Product and culture consulting, Research, Automotives, Telecommunication, AutoML tools.\\
	
	4
	& 9 years
	& Research, Data consulting, Media houses.\\
	
	5
	& 23 years data-intensive, \newline 8 years of ML
	& Cloud analytics, Research, Business model consulting, ML framework development and auditing.\\
	\bottomrule
	\end{tabularx}
\end{table*}

The aim of RQ1 is to survey how the ML models cause problems in software systems.
As for RQ2, we aim to find out what kind of a role fault tolerant design patterns have in ensuring the proper functioning of the systems.
RQ3 is for understanding how fault tolerance should be utilized. If we desire to build better systems, knowing whether or not fault tolerance is seen as useful is not enough in itself, but we also need to know how to do it. Whereas appropriate measures could improve the trustworthiness of the systems, inappropriate ones would just add costs, or worse, add to the misbehaviour.
While all the research questions were approached openly regarding the respondents perceptions, RQ3 also studied the concrete propositions presented in Section~\ref{sec:propositions} although we were not limited to these propositions. 

\subsection{Unit of analysis and case selection}

We designed the study as a multi-case design. We asked several respondents about their experiences over the years on several different projects they had been involved in rather than, for example, only the most recent or successful project. 
Thus, our unit of analysis covered several respondents, each respondent covering several different projects.  While the focus on several projects had the risk of abstraction and missing the holistic view, our research questions focus on discovering the existence and prevalence of characteristics stated in the RQs that can be better covered over several projects. Finally, we focused  our enquiry on the respondents' concrete industrial project experiences from their projects rather than their opinions or general knowledge. 

The respondents for the interviews were selected based on existing academy-industry relations. However, each selected respondent was primarily an experienced  practitioner in the fields of machine learning and software engineering with knowledge of software design (see Table~\ref{tab:respondents}). In fact, none of the respondents had, to the best of our knowledge, a strong academic background, although they typically had an academic degree, and two had some work experience in academia. Respectively, each respondent had at least been involved earlier in some aspects of engineering work in the projects rather than only in a managerial role. All of the respondents had been involved in several projects involving AI or machine learning over the years either in different companies or in different projects for different companies in the case of consultants. We treat the actual projects as confidential information but all projects were conducted within industry, rather than being academic prototypes, and the project customers, which were established businesses, included, but were not limited to, Finnish public authorities, large media houses and companies in the banking sector (cf. Table~\ref{tab:respondents}).  As the problem has a direct practical use, it seemed appropriate to gather the information from those who practice. All respondents worked for Finnish companies with international operations at the time of the interviews.

In total, five interviews were conducted. The number of interviews was not predefined but we aimed at theoretical saturation as used by Strauss~\&~Corbin~\cite{Strauss1994}, meaning that in the last interviews no significant new insights emerged contributing to the research questions and study propositions, but the results and their reasoning were similar and confirmatory to earlier interviews. Thus, we decided not to carry out additional interviews. While the number of interviews remained quite small, the interviews covered much larger number of projects, thus reaching theoretical saturation.

\subsection{Data collection}

The data was collected through semi-structured interviews~\cite{runeson2009guidelines}. The interviews included both open and closed questions but the questions and answers were not expected to follow a strict form, structure, or order. The interviews were conducted in October 2020. Interviews were recorded and supplemented with memos written during the interviews. 

A set of interview questions were designed beforehand. The questions -- and the entire interview protocol -- were carried out in two pilot interviews with a researcher as a respondent using a protocol similar to think-aloud protocol in usability testing~\cite{Nielsen1994} in which the  main interviewer asked the questions and the respondent rephrased how they understood the question, what kind of answer would be given, and provided feedback~\cite{Foddy1994}. The interview protocol was refined based on the pilot interviews.

At the beginning of each interview, confidentiality and other practicalities were agreed upon, the key concepts (as in Section~\ref{sec:background}) were introduced, and the respondents were asked to briefly give their background.
The questions started with open questions, which would capture the respondents' personal preferences and experiences, then widening the scope with more specific and closed questions about single techniques in study proposals (see Section~\ref{sec:propositions}) when the techniques were also introduced. In this way, the respondents were given a chance to describe their views freely, without the propositions giving implications of what we had in mind beforehand, thus supplementing our initial thoughts. The structure follows the order of the research questions so that eventually for each technique in the proposals the respondent was asked i) whether the technique makes sense, ii) whether the technique has been used, and iii) whether the technique could be used. Thus, the questions were of an exploratory nature to gather empirical experiences about the proposals allowing the respondents to express their perceptions.

The questions were presented to the respondents as is but remarks made by earlier respondents were used to attain more in-depth insights, when appropriate, as suggested by Runeson~\&~Höst~\cite{runeson2009guidelines}.
A set of slides\footnote{\href{https://drive.google.com/file/d/1wMtK7FvOZoqdOs0Cd8ZErur2ij7xYXbg/view?usp=sharing}{The slides are available in this link.}} containing the interview questions and study proposals (Section~\ref{sec:propositions}) were prepared and shown to the respondents in order to facilitate better communication. The interviews were carried out remotely because of the on-going COVID-19 restrictions. Each interview took roughly two hours.

\subsection{Data analysis}

The data analysis was conducted as a cross-case analysis~\cite{miles1994qualitative}. The data analysis was started after each interview rather than waiting for all the interviews to be finished. The interview notes and recordings were used as the data for the analysis.

First, the data analysis identified and transcribed respondents' experiences as quotations for the predefined research questions, study propositions, and additional findings beyond these. Quotations from respondents were collected in a concept matrix using a spreadsheet. Each concept in the concept matrix occupied its own row, and each respondent their own column (cf. \cite{webster2002analyzing}). The initial concepts were the RQs and propositions for patterns for fault tolerance (see Section~\ref{sec:propositions}. When a new concept (new pattern, important theme, etc.) emerged from the interviews, a new row was added for it. Quotations from the respondents were added to the cell in their column, which corresponded to the concept that the comment was related to. As the analysis progressed, iterations and refinements were done throughout the concept matrix.

Second, after all the interviews had been analysed individually, synthesis across interviews was carried out. Each concept was analysed cross-case by adding summarising columns, providing an overall view of each of them. For example, for a pattern for fault tolerance, columns for overall impression and perceived usefulness, upsides (pros), limitations (cons), suggestions for implementation, and when to use it were added. The quotations from the interviews were, thus, summarised and generalised in the appropriate new columns.

\section{Results}
\label{sec:results}

\subsection{On misbehaviour of ML systems (RQ1)} \label{sec:misbehaviour}

The mentioned kinds of misbehaviour were unexpected \textit{input-output pairs, poor quality of incoming data}, and \textit{decay of the ML model over time}. The first could be considered the simplest kind of erroneous behaviour: with some inputs, the software gives outputs that are inaccurate or just plainly wrong. In the second kind, poor quality means that the input data is somehow broken or incomplete: For example, if multiple input fields are empty or contain null values, the model cannot be expected to give reliable results.
In the third kind, the model itself does not break, but the distribution of the incoming data varies over time, and no longer matches the training data, causing decay. This decay makes the model -- in a sense -- obsolete. As one of the respondents stated, “prediction models predict the past,” meaning that the models base their results on what has happened before. Thus, if the world changes around the model, the results may not be as good as they were before. 
The change could be the result of, for example, a change in demographics of the user base.

The errors were further divided into \textit{individual errors} and \textit{systematic errors}. Individual errors are errors, which just happen sometimes or with certain inputs. Systematic errors mean that the system always works incorrectly, with a considerable amount of inputs, or with certain sub-problems.

All respondents emphasised that the severity of misbehaviour is contextual. The most frequently mentioned indicator of problematic behaviour was how the misbehaviour reveals itself to the user. For example, a medical appliance is to be expected to give the best and most precise results, whereas for a video recommending system it is enough that the user is somewhat satisfied. 
In the example, there are dire consequences the user of a medical system -- i.e. the patient -- if the system performs poorly, and thus, the tolerance for error is smaller than with the video recommending system. However, the systems must give outputs that are good enough. 
If the users are not satisfied with the results, they will not come back, and if the content managers of the company do not trust the model, the project is deemed a failure.

Misbehaviour is usually the result of \textit{faulty implementation, misuse of the model's results}, or \textit{a poor or buggy model}. The latter is straightforward: the model produces erroneous results. When the model has been implemented into the system in a faulty manner, on the other hand, the model itself gives correct or acceptable results but the software around it breaks the results. One mentioned example of this was an off-by-one-error, in which every user of a platform received the predictions meant for the previous user, thus getting results that were meaningless for them. 
And finally, the misuse of the model's results stems from not understanding correctly what the model does. In this case, the model gets proper inputs, functions correctly, and gives the results that should be expected, but the expectations are wrong or misguided. Thus, the results are used for something that the model was not exactly built for. 

Overall, the respondents raised the challenge that it may be difficult to notice problems. The problems may not be noticed until the system is well into production, but the revenue just will not rise. 
This is particularly true when each feature seems to work separately, but their interaction causes problems.

\subsection{On the role of fault tolerance in ML software (RQ2)}

The need for and role of fault tolerance was deemed to be contextual and varying. As mentioned earlier in Section~\ref{sec:misbehaviour}, systems with direct consequences for the user -- such as medical systems -- are less forgiving than those with none: an error of a medical device can lead even to the death of a patient, while an error in a video recommendation system means -- as one respondent put it -- “looking stupid” and is inconvenient for the service and business. An addition to this -- mentioned by two respondents -- was the role of the human-in-the-loop: Until now, humans have been very much included in the decision-making process, but very recently, more and more critical parts of business have given to automated systems to handle. This raises the value of fault tolerance, as humans are not there to fix every problem immediately. On the other hand, one respondent noted that their system may register a user as a man and a woman at the same time if they are interested in both sports and fashion, and this has not caused significant problems in their prediction model. 
Thus, in some cases, the fault activating might not just be that dangerous.

However, according to the respondents, the fluctuation in the role of fault tolerance is not only about what the system is for, but also about the discipline. The field of ML systems was seen as commercially quite immature -- on both the side of the producer and the buyer -- lacking in established frameworks for developing and maintaining ML systems as complete products. This was discussed a great deal by the respondents. 
The lack of established patterns and frameworks for developing, maintaining, and producing ML-based systems was seen as something that leads to varied and often insufficient practices, as everyone is struggling to come up with their own approaches. In addition, the customers often do not know enough of the field to demand or desire fault tolerance or other kind of safety measures. Quite the opposite: it was stated in the interviews that customers may even become intimidated if fault tolerance, quality assurance, or responsibility are brought up too early in the design phases, even if this is needed when going into production. This was believed to be due to the fact that the field is not yet commercially mature, and the projects are not treated like complete software products with proper requirements from the start.

According to the respondents, in addition to commercial immaturity, problems may arise from the fact that many data scientists come from a purely academic background. For this reason, the data scientist may lack knowledge of the commercial discipline of testing frameworks, continuous delivery, and other common software engineering practices, despite the fact that they often have to implement a great deal of traditional code around the model. This way of working is not alleviated by the perceived gap between data scientists and software engineers. 
Software engineers may not want to participate in the ML side of things, or even believe they cannot participate, while data scientists may not realise that they are dealing with traditional software around the model. In the resulting circumstances, the data scientists and the software engineers may not notice that they could benefit from each other.

\subsection{Patterns used as fault tolerance (RQ3)}

In this section, we present what the respondents thought about the fault tolerance solutions as presented in the study propositions in Section~\ref{sec:propositions}, along with the additional patterns the respondents proposed themselves.

\subsubsection{Input checker}

Input checkers were rarely being used in practice. However, there is use for input checkers, when certain conditions are met.

First of all, hard limits on inputs were seen -- at best -- as an efficient way to prevent poor quality data from entering the model. For example, broken data or data beneath or above some threshold can be filtered out. It may be that the model cannot handle null values, or its results may be unreliable if the user -- for example -- has not watched enough videos for a recommendation.

In addition, business rules and other known problems can be enforced with hard limitations on the inputs' values, forcing the inputs into a certain range. 
Hard limits on inputs can also be used to map exceptionally high or low values to some maxima or minima. For example, in the banking sector, there may not be prior knowledge on how the model acts if the user has an exceptionally high annual income, and it may be a better approach to map to some specific number that is “high enough”, yet known to work.

However, enforcing hard limits excessively was seen as problematic enough that the respondents warned to be cautious. First, the limits may be based on wrong assumptions. If the assumptions are wrong, changes in the model or the incoming data may break the system, as valid inputs will not get through. 
Second, technical aspects could also turn out problematic. One mentioned risk is that the Python programming language is often used in ML projects, and it does not have an inherent type-system. Instead, the type of an input is inferenced during execution time. Thus, the developers may believe they are limiting the inputs, but instead end up breaking or lowering the quality of the inputs.

Checking the inputs for novelty had been used by only one respondent. They, however, considered the novelty check as extremely useful, but not because they saw new kinds of inputs as problematic in themselves, but because of the feedback value of the novel inputs. The recognised novel inputs could be used as training data later on, and also as indicators where previous data was potentially lacking. This is especially valuable, if training data has been difficult to come by. 
Other respondents were somewhat curious about the idea but novelty checking was considered challenging, as recognising such inputs may be difficult.

\subsubsection{Input distribution observing} \label{sec:input_distribution}

Input distribution observing was not one of the original study propositions was but, however, mentioned by every respondent. The statistics of the inputs are measured over time, and deviations in the statistics either alert the developers, or potentially lead to some predefined actions being taken. As mentioned in Section~\ref{sec:misbehaviour}, changes in the world around the model can decay the model's performance. Changes in the distribution of incoming data can be a sign of forthcoming decay, and therefore, something to keep an eye on. If the incoming data starts to differ from the training data of the model, or the data the system had been receiving earlier, this may indicate problems ahead. 

\subsubsection{Output checker}

The respondents considered hard limits on outputs more useful than their counterparts for inputs. Again, business rules or easily confirmable erroneous outputs with direct consequences to users are what set the rules for outputs. For example, business executives might not even approve an autonomous pricing model, which has no limits on how high or low it can value the products. Moreover, the customer should not be able to recognise the erroneous output: For example, it should probably not be possible for the system to place two passengers on the same seat, or give negative values for someone's age.

However, beyond the situations mentioned above, the respondents emphasised that hard limits should be avoided. Not only can the limits be based on wrong assumptions, as is the case with input limits, but also, too many strict limitations can also limit the model's capabilities. If one puts too many limits on the results the model can give, they might only get the results they want, and not the ones they should get.

Using another ML model as an output checker raised interest in the respondents. The model could, for example, predict the trustworthiness of the output, and the output could be accompanied with information that the output may not be very certain. 
However, using another ML model has significant drawbacks in terms of work efficiency, since the checker model should also be validated, and it might be difficult to train in the first place. The training and test data may be hard to acquire, and the acquisition may require breaking the model which is being checked, as the training of the checker model requires known erroneous results for desired outputs.

One respondent presented a way to monitor outputs through \textit{live confirmation}. Basically, if there is a way to gather new, certainly trustworthy data, this data can be used to periodically evaluate whether or not the model is still relevant. For example, the respondent mentioned that they had a way of acquiring some correct information through having registered users along with unregistered ones. In this way, they could occasionally have the model make predictions for registered users, and then compare the predictions to the real values, acquired from their user profile.

\subsubsection{Output distribution observing}

Our study proposal of comparing outputs to historical data was not triumphant when it concerned single outputs. Instead, monitoring the distribution of outputs in a manner similar to inputs in Section~\ref{sec:input_distribution} is something that the respondents mentioned frequently. As with inputs, the distribution of outputs is expected to remain more or less the same. For example, an autonomous application for granting loans should not suddenly approve 100\% of the applications, if the approval rate was previously 50\%. 
Comparing the distributions over time was considered something that should be done more often than what is currently being done. According to the respondents, at the moment, there is a risk that due to the immaturity of the field, developers become too easily satisfied if the outputs match the expected ones once, and forget that the outputs should probably match time after time.

The execution environment often defines whether or not distribution observing is possible when the model has been deployed. It should be remembered that not every ML system is executed in the cloud or another computation-rich environment, where all the data is acquirable by the developers. They can also be, for example, executed on someone's personal device, from where the output distributions may not be within the reach of the developers.

\subsubsection{Model observers}

Measuring the resource consumption of the ML model was mostly disregarded as a tool for fault tolerance for a ML system, but was considered more as a development tool to indicate non-optimal solutions when building an ML model. Alternatively, model observers should be used as something to monitor the execution environment instead of the model and are useful as such considering the development process as a whole. However, the respondents left a side-door ever so slightly ajar in the case of self-learning systems: if the resource consumption metrics started to get worse, the model could learn in a non-optimal way. 

As for activation observers, the experiences were thin. Only one respondent had -- at least knowingly -- used a technique similar to the depicted one. According to them, the inner workings are usually observed by a high-level library, which does not necessarily state what it is observing exactly. 
Another respondent had used an \textit{ad hoc} visualisation of the inner workings of a model in a somewhat similar manner. The rest of the respondents showed interest in the idea, even though they were not aware of such an approach being used.

However, activation observers also have drawbacks. First, the respondent who had used activation observers considered  them labour-intensive. Observing the activations would require knowing specific details about the structure of the model and also having access to the information for a single node's activations. This may not be possible later on in the development process because data scientists often do not build the models as APIs in such a way that would provide the information by merely making function calls.
Also, one respondent considered what should actually be monitored to detect errors instead of design flaws. A surprising node activation does indicate that there is an exceptional situation, but not necessarily why it is exceptional. It might just be saying that the used model is too big. 
On the other hand, rarely activating nodes can also be seen as faults in the model.

\subsubsection{Redundant models}

Having multiple divergent models as recovery blocks to hand the inputs over to was seen as somewhat useful as a fall-over approach in case the main model not give any outputs, or if it was possible to detect erroneous outputs. However, when presented with the idea of having redundant models, the respondents were more inclined to have the system decide the used model based on prior knowledge of the problem, instead of just reacting to bad outputs. The design should be organised in such a way, that certain kinds of inputs would be given to certain models. For example, if two different models perform better on two different kinds of image encoding, a system that gets both kinds as inputs could always give them to the better suited one.
This suggested variant of distributing the inputs to different models based on prior knowledge will later be referred to as \textit{input switch}.

When presenting the idea of redundancy, the respondents often mentioned the \textit{multi-armed bandit models} approach (cf. \cite{lattimore2020bandit}), which was explicitly preferred by two respondents over other approaches to redundancy. Essentially, multiple models compete against each other in these “bandit models”. Inputs are passed on to the models with just some distribution, and over time, the best-performing models are given a greater portion of the inputs, while the worst-performing models essentially fade away. For example, if a user clicks a recommendation, the model is seen as successful, and the frequency of inputs directed to that model can be increased. 
The problem of the changing world can also be addressed by bandit models. Even if one of the models goes down to receive only 1\% of all the inputs, it may be a good idea to leave it in the mix for a while, as in few weeks, the initially poorly performing one might be the one taking the biggest share of inputs. As one respondent put it: “The world changes, and we know that”. 
In addition, introducing new models to the system was found to be safe in bandit models, as the new model can be added amongst the others with a small share of inputs. If the model turns out to be a good one, its share will rise autonomously over time, and if bad, it will fade away. 
Also, a combination of bandit models and some predefined rules, with certain inputs given to certain models, and all the models competing for the rest, was suggested.

Voting was considered as a fairly well-known -- although not overly common -- method, as only one respondent had not seen it being used. The respondents were also mostly satisfied with the voting results they had seen. However, most often the voting was done as a black-box model with a single input and single output, with no interfaces to access the model federation inside. 
A respondent suggested that voting could be enhanced with a better mixture of data science and software engineering skills. Instead of making monolithic models with no transparency, there are situations in which problems could be split into atomic sub-problems, each of which could have a specialised model, all of which would then vote on the outcome. For example, for age estimation, there could be one model that bases its estimation on the person's hair, another for eyes, and a third one for the mouth, and the age would be referenced based on each one. 

Overall, approaches relying on redundancy were seen as something that should be the aim of many projects. When the project becomes advanced enough, and the initial model becomes good enough, the model can be crowned as the “champion” model of the project. New versions should not replace the champion model outright, but instead prove themselves to be better over time. In this way, the project can gather a collection of good quality models intrinsically, and this is something many big companies do in their projects. 
The goal of the system, however, is important. If the goal of the system is to make money, then the models can be used to compete with each other over a longer period of time to find those that make the most money at every given time. On the other hand, medical devices do not have the luxury of time, but should give the best results immediately. Thus, optimising one, fixed approach to its limits could be a better way to go.

Cost is a challenge with all approaches relying on redundancy. Implementing redundancy always requires money and time, which many companies -- especially new ones -- cannot afford. Even maintaining and optimising one model might be a task that gets never finished.
Two respondents mentioned that AutoML tools could decrease the costs. The tools could do the heavy coding for the developers, train numerous different models, and then hand over the best ones, which could then be used as redundant models. 

Recovery blocks consisting of duplicates of the same model were widely disregarded as being more suitable for handling problems in the deployment environment than in the model itself. Thus, they fall outside the scope of this paper.

\subsubsection{Fall-over options}

Over the course of the interviews, \textit{fall-over procedures} were mentioned by the respondents. Essentially, a fall-over means what to do when an error is detected. The recovery blocks of the previous subsection fall into this category as well: when an error is detected, the input is handed over to another model which acts as a fall-over component.

The simplest kind of fall-over action to be taken is to alert the user or developer. Errors and problematic inputs are flagged, and the user or developer is informed about the situation. Then, it is up to the human to decide, what is to come of this. The developer may decide to retrain the model, whereas a common user may contact the administrator of the system. Thus, the user can play a part in the quality control of the system. This, of course, requires that the user knows they are dealing with an intelligent system.

In the case of problematic inputs or erroneous outputs, the results can be ignored or filtered out, meaning that if an erroneous result is given -- e.g., a negative age -- it is just not used. 
This is, of course, only an option if the rest of the system can function without this information.

Also, one respondent had applied a  default output. This means that if no acceptable result is given by the model, some predefined, simple output is given to the user. For example, a recommendation system of an online media platform should not output nothing, as that would leave the front page empty. Instead, for example, the most popular media can be used. 

\section{Discussion}
\label{sec:discussion}

Considering the whimsical nature of ML model misbehaviour, combined with the relative immaturity of the field of ML system development, better understanding and patterns for developing and maintaining these systems are called for. This includes patterns for fault tolerance. When it is hard to know, whether or not the model is actually working when it is in operation, extra measures are needed to keep the systems dependable. While familiarisation with the skills and good practices of software engineering certainly helps, the nature of ML itself must be taken to account as well. The frameworks and solutions are evolving as we speak, and the work is not yet done, but we believe we have provided conceptualisations for misbehaviour, elaborated the need and role of fault tolerance, and identified useful patterns to address different kinds of misbehaviour in ML software.

\begin{siderules}
Key findings are:
\begin{itemize}
    \item ML system can provide poor results if the inputs are of poor quality, the input-output-pairs do not match, or the input distribution drifts. This can be caused by a buggy model, faulty deployment, changes in user base, or misuse of the models results.
    \item Interest in fault tolerance is rising but its overall role and frameworks for it are still developing.
    \item Some patterns for fault tolerance can be -- and already are -- used to tackle the problems caused by the ML model in the system, despite the field still developing.
\end{itemize}
\end{siderules}

\subsection{Discussion on misbehaviour (RQ1)}

\begin{figure}[t]
    \centering
    \includegraphics[width=\columnwidth]{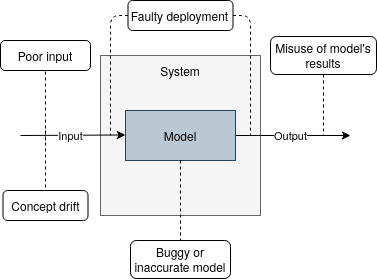}
    \caption{Concepts of misbehaviour and their placement in relation to the \newline system.}
    \label{fig:misbehaviour}
\end{figure}

For misbehaviour (RQ1), we generalised a set of concepts occurring in ML systems as described in Section~\ref{sec:misbehaviour} and illustrated in Figure~\ref{fig:misbehaviour}. The concepts aim to place the misbehaviour in the context of ML system architecture. The concepts themselves are quite general but their implications for different kinds of systems are so vastly different that it would be ill-advised to try to specify which deserves most attention from the developers. For example, one could argue that systematic errors are always worse than individual ones, but even this might be an overstatement if the systematic error does not really reveal itself to the user in a meaningful way. The severity of an extremely rare or invisible systematic error may not be any worse than an extreme individual error. Thus, the severity depends on factors, such as the context and task of the system.

Considering this, the developers' awareness of their system must be emphasised: if the context and task of the system basically define the key cases of misbehaviour, the developers must take this into consideration. Not only that, but it should also show in the system itself. For example, if model decay (also known in the literature as the \emph{concept drift} \cite{tsymbal2004problem}) is recognised as the key threat to a video recommendation system, the choices in design should aim to mitigate that instead of trying to catch every last poor recommendation. In this case, this could mean implementing redundant models as a multi-armed bandit instead of the models voting on every output. The opposite could be true for a brain tumour detection system, as human brains should be quite similar to each other, but as many erroneous outputs should be prevented as possible.

\subsection{Discussion on the role of fault tolerance (RQ2)}

Just as with misbehaviour, also the need for and role of fault tolerance (RQ2) was deemed to be highly contextual and, thus, varying. Not only that, but the relative immaturity of the field and lack of best practices make it difficult to conclusively assess what the role should be. Thus, the problem is tightly intertwined with the development of the discipline. We believe that clearer, proven patterns for risk assessment, model deployment, and system maintenance would tighten the gap between software engineers and data scientists, greatly benefit developers as well as customers -- and actually establish the role of fault tolerance for ML systems.

Typically, the respondents expressed that they would prefer that ML software projects were treated more like software projects and fault tolerance and dependability received more attention. This suggests that the respondents are not satisfied with how dependability is taken into account in the projects. ML software projects become bigger and more business-critical, and their trustworthiness should be addressed all the way through the project, even if the role of fault tolerance is still undeveloped in the process.

\subsection{Discussion on the patterns for fault tolerance (RQ3)}

\begin{table*}[t!]
\small
\caption{A summary of patterns.}
\label{tab:summary} \centering
	\def\arraystretch{1.5}																					
	\begin{tabularx}{\textwidth}{>{\raggedright\arraybackslash}p{1.8cm} >{\raggedright\arraybackslash}p{2cm} R R R}
	
	\toprule

	\textbf{Pattern} & \textbf{Variant}& \textbf{Pros} & \textbf{Cons} & \textbf{When to use}\\
	\hline
	
	\textbf{Input checker}
	& Hard limits
	& Efficient in enforcing business rules and preventing poor quality data entering the model. Computationally light. Low cost.
	& Requires very specific knowledge of the model. May solve only small problems.
	& When accuracy of single outputs is vital, and holey or out-of-range inputs cause problems.\\
	
	& Novelty inputs
	& Shows holes in -- and can be later utilized as --  training data.
	& Difficult to say when a novel input is a problem.
	& When training data is difficult to gather otherwise.\\
	
	\textbf{Input distribution observer}
	&
	& Indicates changes in operations environment, and possible need of retraining.
	& Does not prevent single errors from happening.
	& In naturally evolving or changing input distributions. When input sources are prone to problems.\\
	
	\textbf{Output checker}
	& Hard limits
	& Efficient in enforcing business and safety rules, and in spotting broken outputs. Computationally light. Low cost.
	& Requires very specific knowledge of the domain and system. Careless use leads to limiting results.
	& When business rules or safety regulations dictate a range of acceptable results, or unacceptable outputs can otherwise be recognised for certain.\\
	
	\textbf{Output distribution observer}
	&
	& Indicates changes in operations environment, and possible need of retraining.
	& Does not prevent single errors from happening.
	& In naturally evolving or changing input distributions. When input sources are prone to problems.\\ 
	
	\textbf{Model observers}
	& Resource consumption
	& Could indicate unoptimal development in a continuously learning model.
	& Better suited for development phase and monitoring HW problems.
	& When testing the system after model deployment.\\
	
	& Activation observers
	& Has potential in spotting erroneous input-output pairs.
	& Knowledge claims based on our data cannot be made.
	&\\
	
	\textbf{Redundant models}
    & Recovery blocks with divergent models
    & Offers potentially effective fall-over possibilities.
    & Requires knowledge on when output is erroneous to be applied efficiently. Requires several models. Computationally heavy.
    & When high dependability is required, and other fall-over solutions are too simple for the problem.\\
    
    & Input switch
    & Allows the best suited model to be used for each input.
    & Requires an enormous amount of knowledge about input space and used models. Requires several models. 
    & When inputs can be in several forms or types, or inputs contain several sub-problems.\\
    
    & Multi-armed bandit
    & Raises tolerance against changes in data distribution. Allows safer introduction of new models.
    & Requires several models.
    & In naturally evolving or changing input spaces. If product maturity has introduced several iterations of models.\\
    
    & Voting
    & Potential to eliminate surprising input-output pairs.
    & Difficult to implement. Requires several models. Computationally heavy.
    & When sufficient data science skills are present in the project.\\
    
	\textbf{Fall-over options}
    &
    & Allow simpler, more predictable outcomes when errors are detected. Either brings in the user or developer to solve the problem, or handles situations consistently. Low cost.
    & Results may not be as sophisticated as with finer solutions or models.
    & Often, if not always, as the last resort.\\
	\bottomrule
	\bigskip 
	\bigskip
	\end{tabularx}
\end{table*}
\normalfont 

The different patterns of fault tolerance (RQ3) have different capabilities to address each type of misbehaviour. The refined patterns based on the study proposals that were considered applicable in the case study along with their pros, cons and when to use are summarised in Table~\ref{tab:summary}. 

Hard limits on inputs are efficient in preventing unwanted input-output pairs and poor-quality inputs from entering the model -- \textit{if} you know what you are looking for. Not all erroneous inputs are actually that harmful. This, combined with the fact that ML models that go into production are usually highly accurate with inputs similar to those in their training data, makes it often tedious to find the malign inputs. Moreover, hard limits do not really work to prevent problems caused by drift in input distribution, as the change in distribution does not necessarily mean changes in single inputs in the sense of breaking the limits. Thus, hard limits are usually recommended for limiting inputs to obey business rules, grouping rare inputs into one through, for example, limiting very big or very small inputs to predefined maxima, and clear-cut apparent cases of poor-quality inputs.

Novelty input checker, in turn, could help in observing whether a change in input distribution is happening. However, this is only true if the drift in the distribution brings in formerly unseen inputs, instead of drifting within the same limits. Finding new kinds of inputs is better suited for finding data that was missing in the training data and to improve it -- or potentially find new kinds of ways the inputs can be corrupted. Thus, compared with hard limits, novelty input checker has potential to be a more constructive tool.

The drift in input data distribution can be observed knowingly, and thus, shield the model against possible decay. Not only that, but with systems working within naturally changing environments, like possibly evolving user demographics, it is highly recommended and common, based on the interviews. In addition, skewing statistics can indicate problems in input sources. For example, observing the development of average, standard deviation, and other statistical indicators of the inputs over a batch or a certain period of time, and how they move away from their original values can indicate an upcoming prolapse in the performance of the model. This pattern was brought up by every respondent, even though it was not in the study propositions. This heavily indicates that the initial propositions did not present a full taxonomy.

Limiting outputs with hard limits has similar qualities to limiting inputs. They cannot indicate much about changes happening in the data. Instead, they can enforce business rules effectively, as well as filtering out absurd results: an autonomous system probably should not be able to grant a loan of millions of euros, nor can anyone be less than zero years old. In addition, safety rules can be enforced using hard limits, for example, by preventing an autonomous controller of a machine from surpassing safe levels of torque or heat. Also, outputs filled with missing values, or other indications of clear errors can be monitored. However, one should approach hard limits with caution: having too many limitations on outputs can actually corrode the model, as too strict limitation can actually narrow down the results to exactly what was initially expected, thus harming the generalising possibilities.

Changes in the distribution of outputs can also be a sign of the world changing around the model, similarly to input distribution. According to the interviews, this is a very useful tool, if you can access the output data.

As for redundancy, recovery blocks with divergent models seems useful as a fall-over procedure if there are ways to recognise undesired outputs. As such, it should pair well with output checkers. However, it is costly, and deemed heavy, thus limiting its usefulness. Deciding the used model beforehand by some rules based on knowledge of the problem at hand, and the models' capabilities to handle different sub-problems is another way to organise redundancy. This would limit problems related to unexpected input-output pairs, as the used model should probably be chosen based on the highest accuracy of the specific kinds of inputs. This could also mitigate problems with data distribution drifts, as long as the distribution mostly stays within the strong areas of the models. However, the extensive knowledge required to implement such a system makes large-scale usage impractical. It is probably recommendable to limit the usage to clear cases, such as when a certain model performs better when some input values are absent, or if a certain model performs well with specific sub-problems.

Bandit models with constantly adjusting weights is an efficient way of battling drift in input data distribution, as the best-performing models always gain more ground. In this way, the most suitable models for the current state of the environment overtake the decaying ones. Also, the possibility of adding new models into the federation mitigates the risks in two out of three of the possible sources of misbehaviour: adding a buggy model or accidentally implementing the new model poorly does not immediately break the whole system, as the new model gets suffocated by the better-performing models. However, bandit models are also costly, as is any other design with multiple models. Also, using bandit models requires some metric to be optimised: there has to be some way of measuring the success of each model. This comes naturally to some systems, for example, by registering whether or not the user clicks a recommended product or even buys it, but might not be easy or even possible for some. In addition, compared with choosing the model used based on prior knowledge, adjusting the weights takes time. If the metrics can be measured, bandit models, possibly combined with some predefined rules, have major advantages for tolerating changes in the world around the models, without having to have extensive knowledge of the models' strengths beforehand.

Voting is complicated. Multi-agent models and other ensemble models have successfully used different forms of voting to uplift their accuracy. However, this is often the inner workings of a single, monolithic model, instead of a federation of models with individual inputs and outputs. At best, the models cover each other's weaknesses, and, through voting, mitigate the amount of undesired outputs. Making sure this actually happens instead of amplifying the shortcomings, however, requires skills in data science, and can be a tedious task, not to mention -- again -- the cost of developing multiple models. A federation of models voting on the output is probably well-suited to raising the accuracy of the system, i.e., reducing surprising input-output pairs, and suppressing broken models as they get outvoted, but seems like a difficult and costly approach if the project lacks data science skills.

Observing a model's inner workings through node activations is interesting, and its inclusion in some high-level libraries suggests it has potential. Alas, the details remained so scarce that we dare not make any more detailed suggestions on its usage without further research on the subject. Other model observers, as well as recovery blocks with duplicating models, are useful in their own right, but not so much in the context of this study.

Of course, patterns can be applied simultaneously. Some patterns overlap, and may not work together that well. For example, implementing multiple forms of redundancy sounds like a tedious and costly task, if it can be done in the first place; implementing voting models and recovery blocks in the same system sounds very difficult, as they both involve dealing with multiple, if not all deployed models in a very different manner. However, not all patterns overlap or even touch the same components. It would make sense that the same system had one pattern for inputs, another for outputs, and some redundancy, if each of them is seen as useful and implementable in the system. Moreover, some sort of fall-over option should probably be implemented in every system.

Taking the idea a step further, different patterns could be used to support each other, as different patterns are effective against different kinds of misbehaviour. If concept drift is deemed to be the biggest threat to the system, it does not sound unreasonable to observe the input and output distributions and implement a multi-armed bandit to supplement each other in the task. Likewise, if input and output checkers and voting models can be implemented, they could potentially supplement each other in an effort to minimise the amount of erroneous results.

Not all forms of redundancy rule each other out. An interesting idea would be having certain cases for which predefined rules would steer specific kinds of inputs to specialised models, while dealing with most of the inputs using bandit models or voting. In this way, the system can benefit from prior, explicit knowledge, as well as the implicit knowledge provided by the whole orchestra.

Many of the patterns bear a resemblance to patterns for fault tolerance in traditional software \cite{knight2012fundamentals}. They usually do, however, have a ML-specific twist to them. For example, inputs and outputs are not necessarily limited because the values are known to be erroneous or dangerous, but because the values are known to be problematic in relation to certain business rules, or too rare in the training data to be trusted. Also, some have not risen directly from traditional concepts, and are very much based on the basics of ML. For example, observing inputs for changes in distributions or possible novel inputs are based on the threats of concept drift and incomplete training data. Some have traditional \emph{and} ML-specific implications: using bandit models not only establishes which model functions best in the context when implemented, but also protects the system from concept drift, as the most suitable model for the moment will emerge on top.

Overall, we have presented a set of patterns for the fault tolerance of ML systems (Table~\ref{tab:summary}). To our knowledge, this is the first attempt to synthesise such a set. The patterns are gathered from experiences of experienced practitioners, refined from the initial model (see Section~\ref{sec:propositions}) through adding, detailing, and rejecting patterns. Thus, scarce, anecdotal information from earlier research has been clarified in context, while supplementing it with new patterns and new information on how to utilise patterns inspired by traditional fault tolerance. The patterns have different strengths and shortcomings, while each is also more suitable for addressing certain problems better than others. Thus, even though the field and role of fault tolerance is still immature, and the set may not be complete, we believe the patterns to be useful for addressing different, actual problems when building ML systems in different contexts.

\section{Validity}
\label{sec:validity}

We base our validity discussion on the work of Shadish \textit{et al.}~\cite{shadish2002experimental}. Since we carried out a case study as a form of qualitative study and did not make statistical conclusions, we focus only on the internal, construct, and external validity.

Internal validity means the validity of causality: do the treatment and the outcome actually reflect the causality between them? In this paper, this would mean whether or not the presented architectural designs actually promote fault tolerance. The respondents were asked to elaborate on what kind of misbehaviour they have witnessed pertaining to ML software based on the general description of faults and errors, and asked how these problems were addressed in the software. Furthermore, considering the case propositions, the respondents were not only asked whether they saw the propositions' potential, but also whether they had actually used something similar themselves, or would. Also, the respondents were not hesitant to reject propositions or suggest alterations.

Construct validity means the validity of conceptualisation and theoretical generalisation, i.e., are the concepts properly defined and understood? A prevalent problem in interview-based studies is whether the respondents understand what the interviewer meant, and vice versa. To tackle this problem, the central concepts were explained to the respondents at the beginning of every interview, the slides were shown to help everyone to follow the interview, and the descriptions of case propositions were backed up by diagrams of the proposed design shown in the slides. As several respondents gave similar answers and similar suggestions, and also outright rejected some propositions as not relevant to the described situation, we believe the concepts were clear for the respondents. On the other hand, to mitigate the threat that we as researchers misunderstood the respondents, two interviewers were always present at each interview, the interviews were recorded and the recordings were used in the analysis, forming a chain of evidence, and we gave the respondents an opportunity to review a draft of this manuscript. Another threat to  construct validity is that many of the proposed cases stem from experiences in traditional software engineering; thus, they might be out of place in the world of ML systems. However, the respondents were ready to reject propositions they found not to be useful concepts in the field at hand.

External validity refers to the generalisability of the results. A threat to external validity in qualitative studies pertains to the selection of the respondents for the study.
First, the respondents might share similar views because of commonalities in their background and projects they have participated in. Second, all the respondents work for organisations in Finland, which gives a geographically and culturally limited selection. Third, the number of respondents was five, which can be considered quite low. However, the respondents have different educational backgrounds, and have worked for a range of organisations, for example, in start-ups, big media companies, and -- to a smaller extent -- academia. The organisations and their business environments are actually relatively international, at least focusing on European markets. Also, despite the number of respondents, their responses were not limited to a single project, but they were allowed to draw from their experiences as a whole. This could be seen in the fact that respondents used a wide variety of examples in their responses, and often suggested that some case propositions might be useful in contexts they usually did not use as examples. The variety in the examples could suggest a good variety in the applicability. Of course, this does not mean that the suggested patterns would work in all cases, nor does it mean that they cover all the possibilities even in the explicitly mentioned situations. That is, the results as the outcome of a case study are not statistically representative or a necessarily full taxonomy. However, as the responses saturated along with the interviews, we believe we have gathered a fairly good body of knowledge on the current state of the field. 

\section{Conclusions}
\label{sec:conclusions}

We have presented a case study on the fault tolerance and misbehaviour of software systems utilising ML. The results were drawn from five semi-structured interviews with experienced architects from the industry. 
The need for established frameworks for \textit{ML software engineering} is on the rise as ML models are being deployed in more and more business-critical and autonomous environments. This is not limited to fault tolerance, but also other phases in the life-cycle of ML software. The more direct consequences the system has for the life of the user, the more care should be taken when ensuring the dependability of the system. However, as the field is still quite immature, fault tolerance is still trying to find its place. Some patterns are used in the industry, but practitioners and customers often do not consider the ML system as a complete software engineering project with a need for fault tolerance.
Regarding misbehaviour and fault tolerance, our study also emphasised the consideration of ML system's context. 

The currently applied patterns for the fault tolerance of ML systems can be roughly divided into input and output checking, data distribution observing, redundancy, fall-over options, and their sub-patterns. Input and output checking and autonomous orchestration of inputs to the most suitable redundant models were the most used designs to detect errors and prevent them from propagating. Voting is also a useful solution for this. Redundant models as recovery blocks, along with default outputs and other fall-over options, are used for error recovery when an error is detected. Data distribution observers and bandit models are highly regarded as tools to keep the models relevant, even in the face of the world changing around the systems. Model observers deserve further research.

These are hardly the only ways to build fault-tolerant ML systems, and this is just the first attempt to gather this information. The field is young, and as frameworks for the development of these systems are tuned further, fault-tolerance patterns will no doubt find their place and have their share of development.

\section*{Acknowledgements}
This work was funded by local authorities under grant agreement ITEA-2019-18022-IVVES of ITEA3 programme and the AIGA project.
We acknowledge the help of Elina Kettunen in writing the manuscript and Vlad Stirbu in refining the interview protocol.

\bibliographystyle{elsarticle-num-names}
\bibliography{ref.bib}
\end{document}